\documentclass[twocolumn,showpacs,preprintnumbers]{revtex4}

\usepackage{amsmath}
\usepackage{amssymb}
\usepackage{graphicx}
\usepackage{dcolumn}
\usepackage{bm}

\begin{document}

\title{Considerable enhancement of the critical current in a superconducting
film by magnetized magnetic strip}

\author{D. Y. Vodolazov}
\email{vodolazov@ipm.sci-nnov.ru}
\author{B. A. Gribkov, S. A. Gusev, A. Yu. Klimov, Yu. N. Nozdrin,
V. V. Rogov}
\author{S. N. Vdovichev}
\email{vdovichev@ipm.sci-nnov.ru}

\affiliation{Institute for Physics of Microstructures, Russian
Academy of Sciences, 603950, Nizhny Novgorod, GSP-105, Russia}

\date{\today}

\begin{abstract}
We show that a magnetic strip on top of a superconducting strip
magnetized in a specified direction may considerably enhance the
critical current in the sample. At fixed magnetization of the
magnet we observed diode effect - the value of the critical
current depends on the direction of the transport current. We
explain these effects by a influence of the nonuniform magnetic
field induced by the magnet on the current distribution in the
superconducting strip. The experiment on a hybrid Nb/Co structure
confirmed the predicted variation of the critical current with a
changing value of magnetization and direction of the transport
current.
\end{abstract}


\maketitle

\section{Introduction}

The best-known and useful property of superconductors is the
ability to carry current without dissipation. Unfortunately, in
type II superconductors the magnetic flux may enter a
superconducting sample in the form of Abrikosov vortices, and
their motion under the influence of the Lorentz force (produced by
transport current) leads to dissipation of the energy
\cite{Tinkham}. There are two ways to prevent this motion: the
inhomogeneities of the sample which pin the vortices
\cite{Campbell} (so called bulk pinning) and/or
surface/geometrical barrier \cite{Bean,Likharev,Fortini} which
does not allow vortices to enter/exit the sample. For both cases
there is a critical current $I_c$ above which the pinning centers
or the surface barrier do not hold vortices any more, so
dissipation starts in the superconductor. For the bulk pinning
case at $I=I_c$ in any point of the sample there is equilibrium
between the Lorentz force $F_L=\Phi_0j/c$ ($\Phi_0$ is the
magnetic flux quantum) and the pinning force $F_p$, and the local
current density is equal to the pinning current density
$j(r)=j_p(r)$. In the case of the surface barrier dissipation
starts when the current density on the surface/edge exceeds the
critical value $j_s$ (for a defectless superconductor $j_s$ is
equal to the Ginzburg-Landau current density $j_{GL}$) and in any
point of the sample the current density has the same sign
\cite{Kupriyanov,Aslamazov,Benkraouda} (the last condition
provides the vortex passage through the sample).

There is a theoretical limit for the critical current - it cannot
be larger than the product of the Ginzburg-Landau current density
and the cross-section $S$ of the sample $I_{c}^{theor}=j_{GL}S$.
In the bulk pinning case vortices become depinned normally at
$j_p\ll j_{GL}$ and $j_p$ usually decreases with an increasing
local magnetic field \cite{Campbell}. In a strip with the
transport current the field is maximal on the edge of the sample
and, hence, the pinning current density is minimal there. The
situation is opposite to the above for the surface barrier
mechanism - the current density is maximal on the edge and minimal
inside the sample at $I=I_c$ \cite{Kupriyanov,Benkraouda}. As a
result, in both situations the real critical current is much
smaller than $I_c^{theor}$.

One interesting way to increase the critical current (besides the
attempts to increase the pinning current density by artificial
pinning centers) is the use of magnetic or superconducting screens
\cite{Genenko1,Genenko2,Genenko3} around a superconducting sample.
The main idea of this method is to make the current distribution
in the sample homogeneous due to screening of the current induced
magnetic field. It was shown theoretically that for the surface
barrier mechanism the critical current may reach the maximal
possible value $I_c^{theor}$ by this method \cite{Genenko1}.
\begin{figure}[hbtp]
\includegraphics[width=0.45\textwidth]{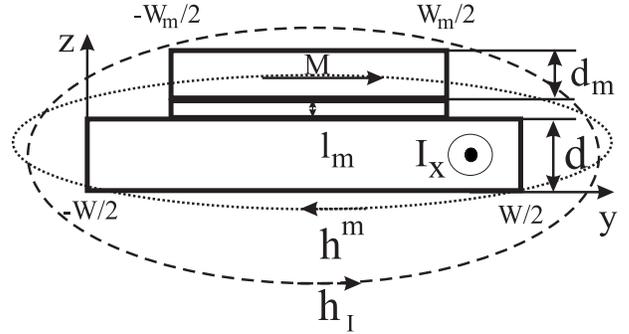}
\caption{Magnetic strip with thickness $d_m$, width $w_m$ and
magnetization $M$ placed on the top of superconducting strip with
thickness $d$ and width $w$. Thickness of the isolating layer is
$l_m$, and dashed and dotted lines show qualitatively the magnetic
field lines from the magnet and the superconducting strip with
transport current $I$ in the X direction.}
\end{figure}

In our paper we propose another method for enhancement of the
critical current by magnetized magnetic materials. We apply a
nonuniform magnetic field induced by a magnetic strip to a
superconducting film with a transport current. The easiest way to
do that is to place the magnet on the top of the superconducting
strip (Fig. 1). It is clear from the figure that, depending on the
relative direction of the magnetization and the transport current,
it lead to a decrease or increase of the total magnetic field
inside the superconducting strip. It may result in smoothing of
the current distribution and enhancement of the critical current.

In the paper we quantitatively study the value of the enhancement
of the critical current for both the bulk pinning and the surface
barrier mechanisms. We show theoretically that for real materials
and realistic parameters it is possible to increase the critical
current several times using this method. We assume in our model
that the magnetic field induced by the current is unable to change
the magnetization of the magnetic material. It is valid in two
cases: i) if this field is smaller than the coercive field of the
ferromagnetic material; ii) if we apply the magnetic field in
parallel to the strip to compensate for the current induced field
and magnetize the magnet. The experiment on a Nb/Co structure with
the parameters being far from optimal gave us an increase in the
critical current by 20 $\%$.
\begin{figure}[hbtp]
\includegraphics[width=0.42\textwidth]{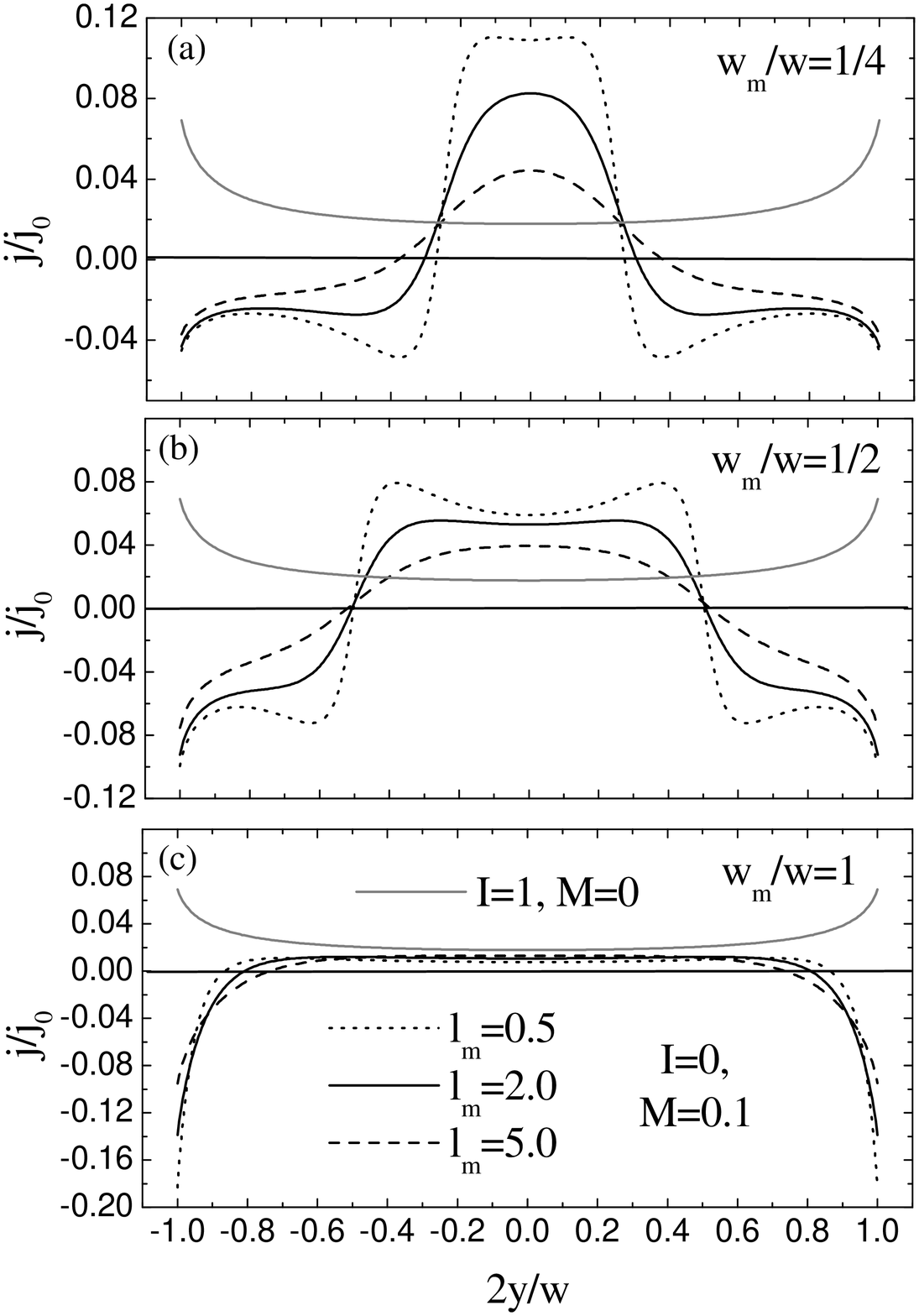}
\caption{Distribution of the current density in a superconducting
strip (w=40, d=1), induced by a magnetic strip (d$_m$=1, M=0.1,
I=0) for different widths w$_m$ and separation distances l$_m$.
Gray curve shows the current distribution in the superconducting
strip with current I=1 and zero magnetization M=0.}
\end{figure}

The paper is organized as follows. In Sec. II and III we study
theoretically the value and the conditions for the enhancement of
critical current in the structure shown in Fig. 1 for two
irreversibility mechanisms - the surface barrier and the bulk pinning,
respectively. In Sec. IV we present the results of our
experiment on a Nb/Co structure, and in Sec. V we discuss the
restrictions and conditions for observing this effect in other
superconducting materials.

\section{Surface barrier mechanism}

First, let us consider the case when the critical current is
determined by the surface barrier effect. As already mentioned
above, the sample is in the critical state when on the edge the
current density reaches the critical value $j_{s}$ and nowhere
inside the sample does $j$ change sign.

To find the critical current for system shown in Fig. 1 we use the
model equation \cite{Kupriyanov,Aslamazov,Benkraouda,Larkin}
\begin{equation}
\frac{dj(y)}{dy}+\frac{d}{2\pi} \int_{-w/2}^{w/2}
\frac{j(y')}{y'-y}dy'=h_z^0-n(y)\Phi_0,
\end{equation}
which describes distribution of current $j(y)$ and vortex density
$n(y)$, averaged over the strip thickness and inter-vortex
distance, in presence of transport current and external uniform
magnetic field $h_z^0$. In the presence of a magnetized magnetic
strip we should add in the right hand side of Eq. (1) the magnetic
field $h_z^m$
\begin{eqnarray}
h_z^m(y)=2M/d(F(y,l_m)-F(y,l_m+d)
\nonumber \\
 -F(y,l_m+d_m)+F(y,d+l_m+d_m))
\end{eqnarray}
\begin{eqnarray}
F(y,a)=-\frac{a}{2}\log\left(\frac{a^2+(y-w_m/2)^2}{a^2+(y+w_m/2)^2}\right)+
\nonumber \\
(w_m/2-y)\arctan\left(\frac{a}{y-w_m/2}\right)+
\nonumber \\
(w_m/2+y)\arctan\left(\frac{a}{y+w_m/2}\right)
\end{eqnarray}
induced by magnetic strip the magnetized in the Y direction (${\bf
M}=(0,\pm M,0)$) and averaged over $d$. In Eq. (1) the distance is
measured in units of the London penetration depth $\lambda$, the
current density is in units $j_0=c\Phi_0/8\pi^2 \lambda^2 \xi$
(where $\xi$ is the coherence length). Magnetic field and
magnetization are scaled in units of $h_c=\Phi_0/2 \pi \xi
\lambda$. In general Eq. (1) is valid for arbitrary thickness $d$
\cite{Kupriyanov,Benkraouda} but one should be careful when
applying to find the critical current of a thick strip with
$d>\lambda$ and a magnet on the top of it. Indeed, for such a
sample the current distribution is strongly nonuniform over the
strip thickness, which leads to a nonuniform force (it is stronger
on the top and weaker at the bottom of a superconducting strip)
acting on the vortex. We may say that in this limit the results
obtained from solution of Eqs. (1,2) should be considered as a
semi-quantitative estimation.
\begin{figure}[hbtp]
\includegraphics[width=0.42\textwidth]{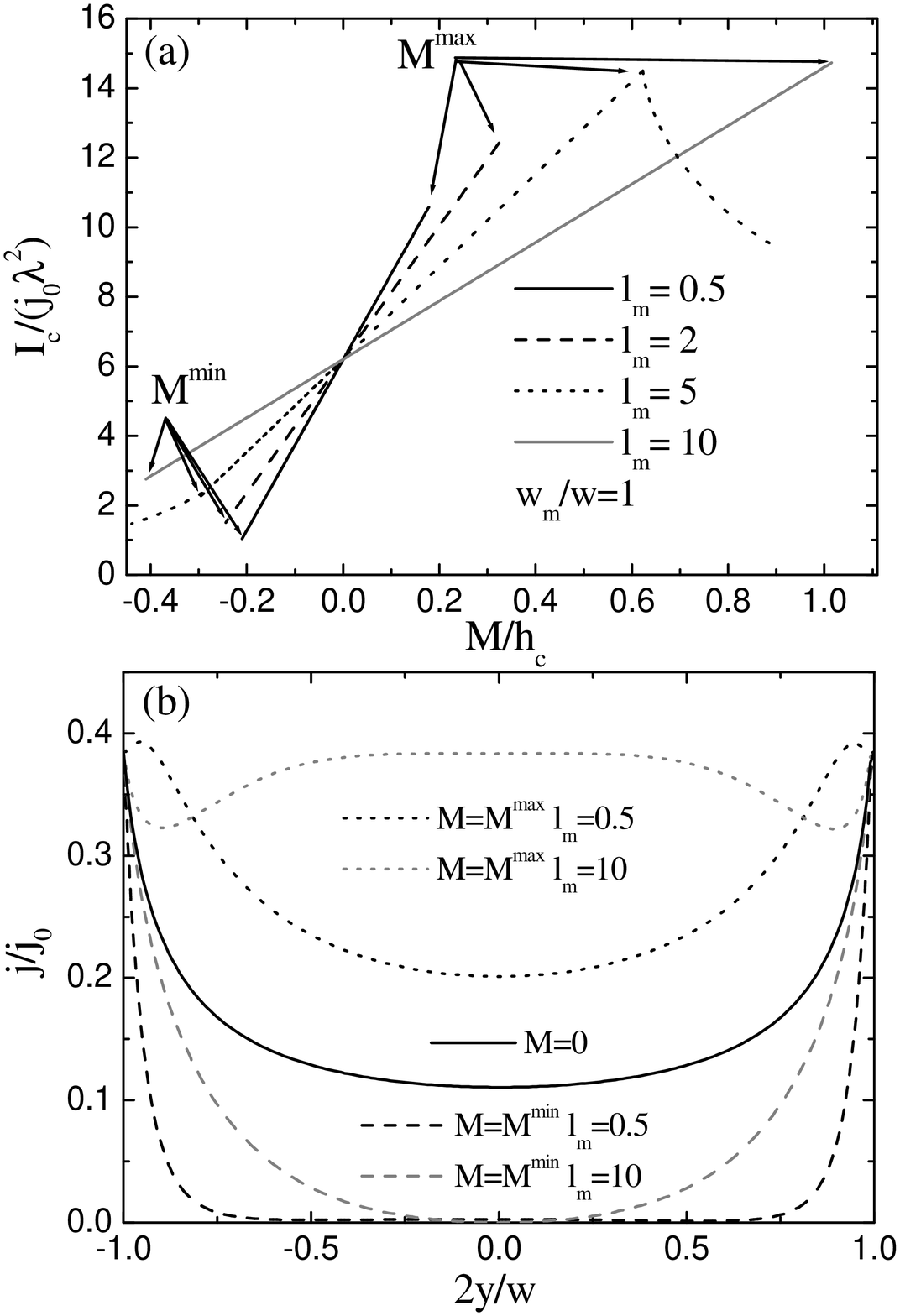}
\caption{Dependence of the critical current (a) in the
superconducting strip with a strong surface barrier effect as a
function of the magnetization of the magnetic strip magnetized in
the Y direction. In the range $(M_{min},M_{max})$ there are no vortices
in the strip at $I<I_c$. In part (b) we plotted the current
distribution at $I=I_c$ for $M=M_{max}$ and $M=M_{min}$.
Parameters of the hybrid system are the same as shown in Fig. 2c. Dotted curves
in figure (a) show the qualitative behavior of $I_c(M)$ at
$M>M_{max}$ and $M<M_{min}$.}
\end{figure}

We consider the case of an applied zero magnetic field in the Z
direction, $h_z^0=0$, assuming that the edge of the sample is
defect-free ($j_s=j_{GL}=\sqrt{4/27}$). In Fig. 2 we plotted the
distributions of the current density induced by a magnetized
magnetic strip (at different $w_m$ and $l_m$) or by a transport
current. It is obvious that there is an optimal (for every
specific ratio $w/\lambda_{eff}$) distance $l_m$ and width $w_m$
at which the enhancement of $I_c$ would be strongest. The reason
is that, if the magnetic strip is very close to and/or narrower
than the superconducting strip the total current distribution
(from the magnet and the transport current) may be more nonuniform
than that from the transport current alone (see Fig. 2). Usually
the enhancement effect is maximal when $w_m \simeq w$ and for our
parameters ($w=40$, $d=1$, $d_m=1$) the thickness of isolating
layer of about $ \simeq 5$ is optimal for reaching $I_c^{theor}$.

In Fig. 3a we presented the dependence of the critical current (in
the X direction) on the sign and value of the magnetization $M$ at
different $l_m$ and $w_m$. It always has a linear dependence on
the $M$, if there are vortices in the sample, with a slope
depending on $d_m$, $l_m$ and $w_m$. At $M=M_{max}$ the critical
current is maximal for the given geometrical parameters of
superconducting and magnetic strips because at $M>M_{max}$ the
vortices start to nucleate somewhere inside the sample rather than
at the edge, (where $j$ is maximal - see Fig. 3b) and $I_c$
decreases. At $M<M_{min}$ vortices appear in the superconducting
sample even at $I<I_c$ and the critical current decreases more
slowly than by the linear law, with a further decreasing $M$. At
the parameters choice  of the superconducting strip it is possible
to increase its critical current by the proposed method more than
2.5 times (it almost reaches the theoretical limit
$I_c^{theor}=j_{GL}wd\simeq 15.4$).
\begin{figure}[hbtp]
\includegraphics[width=0.42\textwidth]{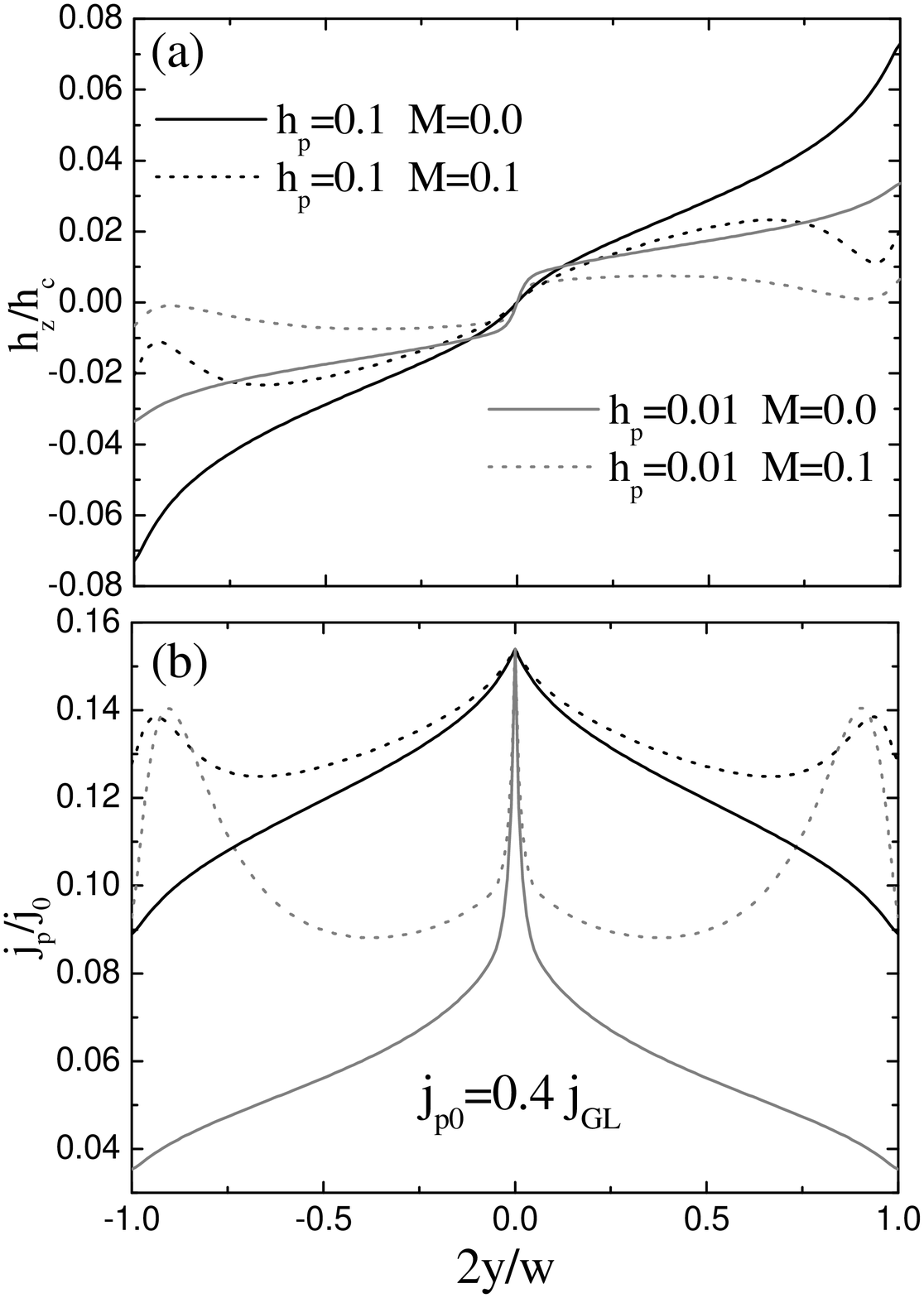}
\caption{Magnetic field (a) and current density distribution (b)
in the superconducting strip for different values of strength of
pinning (see Eq. (5) for other values of $h_p$) and magnetization
$M$. With a decreasing $h_p$ the effect of the magnetic strip
becomes stronger.}
\end{figure}

It is clear that if we change the direction of the current (at
fixed magnetization), we change the direction of variation of $I_c$
(see Fig. 1). We may write that $I_c^{+}(M)=I_c^{-}(-M)$ where
$I_c^{+}(M)$ is the critical current in the X direction and
$I_c^{-}(M)$ is the critical current in the opposite direction.

\section{Bulk pinning mechanism}

In contrast to the previous case, a superconducting strip is
filled up with vortices in the critical state (at $I=I_c$).
Their distribution is determined by the condition that in every point of
the sample the current density is equal to the pinning current
density. The equation for distribution of the magnetic field and
the current density is a kind of the Bio-Savar's law
\begin{equation}
h_z(y)=h_z^0+h_z^m(y)+\frac{d}{2\pi} \int_{-w/2}^{w/2}
\frac{j_p(y')}{y-y'}dy',
\end{equation}
in which the pinning current density depends on the local magnetic field
$j_p(y)=j_p(h_z(y))$. In our numerical calculations we use the
well-known Kim-Anderson model \cite{Campbell,Kim}
\begin{equation}
j_p(h_z)=\frac{j_{p0}}{1+|h_z|/h_p}
\end{equation}
which we inserted in Eq. (4).
\begin{figure}[hbtp]
\includegraphics[width=0.45\textwidth]{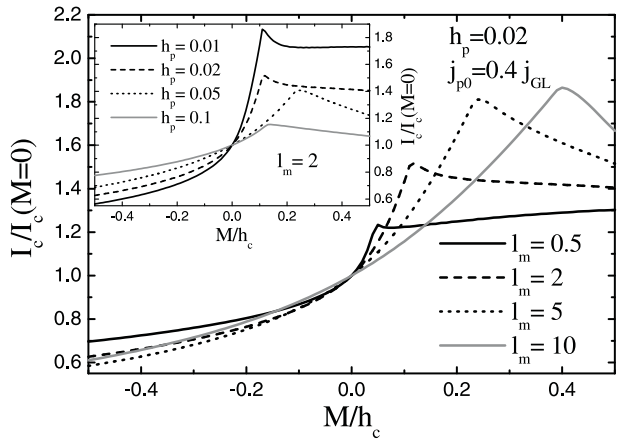}
\caption{Dependence of the critical current (in bulk pinning
model) on the magnetization of the magnetic strip for different
distances $l_m$. The parameters of the strips are the same as in
Fig. 2c. In the inset we show the variation of $I_c(M)$ with a
change in pinning (for different $h_p$).}
\end{figure}

In the framework of model Eq. (5) we may claim that the use of a
magnetic strip could largely increase $I_c$ if the current induced
field $h_I$ is comparable with or larger than $h_p$ on the edge of
the superconducting strip. When the magnetic field induced by the
magnetic strip compensates $h_I$, it causes an increase in the
pinning current density and in the critical current. Fig. 4
illustrates this phenomenon (here we considered only the case
$w_m=w$ which provides maximal enhancement of $I_c$). When we
decrease the field $h_p$ (keeping other parameters constant), the
effect of field compensation becomes more pronounced (see the
inset in Fig. 5) and the critical current approaches a maximum
value for this model: $I_{c}^{max}=j_{p0}wd$. Also, as for the
surface barrier mechanism, there is an optimal distance $l_m$,
where the effect is strongest (see Fig. 5) for all other
parameters being constant.

\section{Experiment}

Actually, the idea of using a magnetized strip to enhance the
critical current in a superconducting strip was originated from
the experimental work \cite{Fraerman}, where the effect of a chain of
ferromagnetic particles on $I_c$ of a superconducting bridge was
studied. The critical current in the X direction increased when the
magnetic particles were magnetized in the Y direction \cite{Fraerman}.
We explain that result by the influence of the magnetic field
induced by the magnetic particles (in the way considered above) and
to check this hypothesis we made an experiment with a simpler geometry.

In this work, for experimental investigation of the effect of the
an inhomogeneous magnetic field of the magnet on the critical
current of a superconducting film, we fabricated a narrow $Nb$
bridge with a Co line positioned under the center of the bridge.
Figures 6(a,b) show an AFM image of the structure under study. The
bridge was characterized by the following parameters: the
thickness $d$ was about 100 nm, the lateral dimension of the
constriction width $w=2 \mu m$ and the length $L=12 \mu m$, the
critical temperature was about $9.2 K$. The ferromagnetic $Co$
strip was obtained by electron lithography \cite{Fraerman2} and
had following dimensions: width $w_m=0.4 \mu m$, length $L=14 \mu
m$ and thickness $d_m=100 nm$. The ferromagnetic and
supercondacting strips were separated by a thin ($l_m=50 nm$)
layer of insulator material (to prevent the proximity effect). The
magnetic state of the $Co$ line was monitored by a Solver scanning
probe microscope at room temperature in `flying` mode. Figure 6(c)
shows an MFM image of the sample. The stripe domain structure of
the $Co$ strip (with the residual magnetization close to zero) is
clearly visible.

The measurements were performed at a temperature $T=4.2 K$ by the
standard four probe method. The dependence of the critical current
(along the X axis) on external uniform magnetic field applied in
the Y direction was measured. Note that the critical current of
the blank Nb bridge (without magnet) is independent of the
external magnetic field in the X and Y direction up to 3 kOe.

Figure 7 shows the results of measuring $I_c(H)$ for the positive
and the negative transport current, respectively. We observed
variation of the critical current when the magnetic strip is
magnetized in the Y direction. There are two effects. First, the
critical current $I_c^{+}(H)$ enhances with an increasing external
magnetic field. Second, there is a strong asymmetry for different
directions of the current, so called diode effect (see
current-voltage characteristic in the inset in Fig. 7). The value
of the diode effect is about 180\% in the dc regime. Actually, the
weak effect was also found when we magnetized the strip in the X
direction (variation of $I_c$ was about $5 \%$). We explain it by
appearance of an uncontrolled components of the magnetic field
induced by nonuniform magnetization distribution of the magnetic
strip (Fig. 6c).

\begin{figure}[hbtp]
\includegraphics[width=0.45\textwidth]{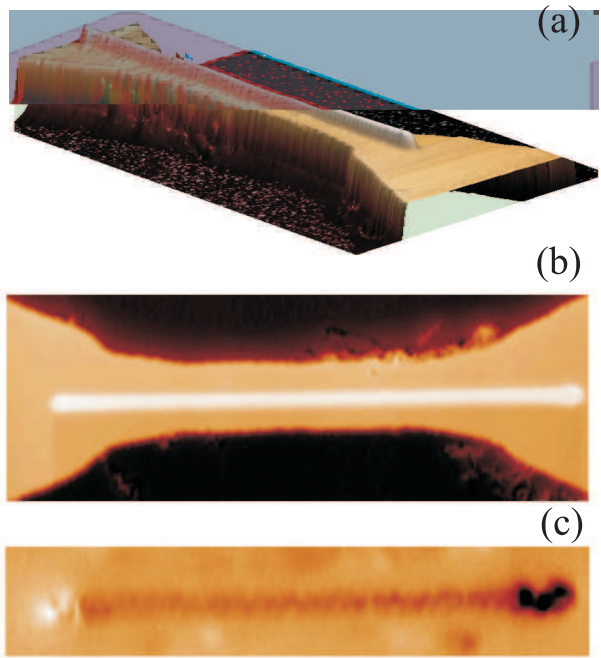}
\caption{3D (a) and 2D (b) AFM images of our niobium bridge with a
cobalt strip on the top. In part (c) we show the MFM image of our
cobalt magnet in demagnetized state.}
\end{figure}

In our experiment we also observed a hysteretic dependence of
the critical current on the magnetic field in both X and Y
directions. It occurs already after the first sweeping up and
down the applied magnetic field. We attribute it with the hysteresis in
the process of magnetization of the cobalt strip. From these
measurements we found a coercive field of our magnetic strip,
$h_{coer}\simeq$ 180 Oe, in the Y direction at T=4.2 K.

\section{Discussion and Conclusion}

The increase in $I_c$ was about $20\%$ for our specific
geometrical parameters (it is close to the value observed in
\cite{Fraerman} for a chain of magnetic particles). If we use the
model of surface barrier and parameters typical for dirty Nb
\cite{Gershenzon} ($\lambda \sim 100-200 nm$, $\xi \sim 20-10 nm
$, h$_c \sim 1600$ Oe) we find the same maximal increase in the
critical current for our hybrid system. For the bulk pinning model
(with $j_{p0}\sim 4\times 10^6 $A/sm$^2$ and $h_p \sim 250$ Oe
found from the best fit to experimental results for $I_c(h_z^0)$)
we found a much smaller theoretical enhancement of the critical
current. These results lead us to believe that the surface barrier
plays an essential role in our experiment, at least at
$h_z^0\simeq 0$. It is in agreement with the results of Ref.
\cite{Gershenzon}, where the importance of the surface barrier
effect was experimentally proved for similar Nb bridges at low
magnetic fields. We are planning to continue our research on the
wide superconducting and magnetic strips to optimize enhancement
of the critical current.

Our experiment shows that it is possible to observe the diode
effect in such a structure and control its value by variation of the applied
magnetic field (Fig. 7). We may theoretically estimate, using the
standard model of viscous motion of vortices and the heat balance
equation, that the diode effect exists for our structure at
frequency $\nu \lesssim 10^5$ Hz of the applied ac current.

\begin{figure}[hbtp]
\includegraphics[width=0.45\textwidth]{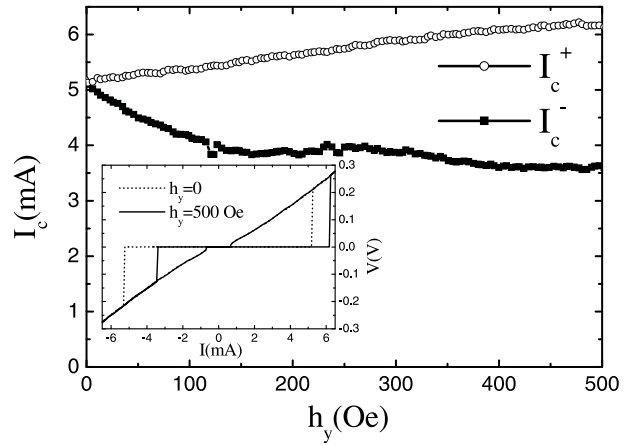}
\caption{Experimental dependence of $I_c^{+}$ (critical
current in the X direction) and $I_c^{-}$ (critical current in the
opposite direction) on the applied magnetic field in the Y direction.
At $h_y=0$ the magnetic strip was demagnetized (remanent
magnetization is equal to zero). In the inset we presented the dc
I-V characteristics of our hybrid system at $h_y=0$
and $h_y=500$ Oe, which show a pronounced diode effect at large
$h_y$.}
\end{figure}

In the theoretical model we neglected the effect of the current
induced magnetic field on magnetization of the magnetic strip. In
our experiment it does not play any role because we applied a
parallel magnetic field to the sample. This field did not affect
the critical current of the sample directly because the parallel
critical field is about $h_cd/2\lambda \sim 3200-6400 Oe$ for our
thin Nb bridge.

In our calculations within the surface barrier model we assumed
that the vortices enter the sample when $j_{edge}=j_{GL}$.
Actually, in a real situation there always are some surface
defects which favor the vortex entrance and diminish the surface
critical current density \cite{Buzdin,Aladyshkin,Vodolazov}. This
does not affect the main result because the current density is
still maximal at the edge and minimal in the center of the strip
with surface defects at $I=I_c$. Enhancement of $I_c$ may be even
larger because $j$ inside the sample may be larger than the
current density at the edge (but cannot exceed $j_{GL}$).

The larger the ratio $w/\lambda_{eff}$, the higher nonuniformity
of the distribution of the current density over the strip width
(see analytical expression in Ref. \cite{Plourde} for an arbitrary
value of $w/\lambda_{eff}$). It means that covering by magnetic
material in order to affect the critical current is more effective
for wide films with $w/\lambda_{eff}\gg 1$.

The situation is more complicated when both the surface barrier and the bulk
pinning play an essential role. Nevertheless, the current distribution
would be still nonuniform \cite{Elistratov} and we expect the
predicted effect to exist.

The materials with a large magnetization which can be used in
experiments are cobalt or iron with $M_{sat}\simeq 1800$ Oe. A
good candidate for observing the predicted effect in the case of
surface barrier mechanism is amorphous MoGe. The pinning current
density may be as small as $10^2$ A/cm$^2$, and the experiment
shows a pronounced surface barrier effect (with $j_s \sim
j_{GL}\sim 10^6$ A/sm$^2$) on the critical current (see Ref.
\cite{Plourde}). For this material $h_c \sim 800$ Oe (with $\xi
\sim 7.5 nm$, $\lambda\sim 560 nm$) and, in dimensionless units
$M_{sat}/h_c \sim 2.2$. It means that the predicted effect may be
easily observed in this material (see Fig. 3a). It is known that
the surface barrier plays important role for YBaCO high
temperature superconductors \cite{Konczykowski,Burlachkov}. In
this material $j_{GL}\sim 10^8 $ A/sm$^2$, $h_c \sim 10^4$ Oe
(with $\xi \sim 2 nm$, $\lambda\sim 200 nm$), $M_{sat}/h_c \sim
0.2$ and one can also observe enhancement of the critical current.
As in our experiment, a parallel magnetic field can be applied to
the hybrid system to magnetize the magnetic material and diminish
the effect of the current induced field (the Y component of the
magnetic field induced by the transport current can be
overestimated as $j_{GL}d/2 \sim h_c/8$ at $d\sim \lambda$ and,
hence, can change the magnetization of the cobalt magnet with the
coercive field of about $180$ Oe for our magnetic strip).

The magnetic field induced by the magnetic strip decays as $\sim
1/r^2$ (at $r \gg w_m$) and the one induced by the superconducting
strip with transport current as $\sim 1/r$ (at $r \gg w$) only.
Actually, it is the most important property of our hybrid system
useful for applications (for a superconducting magnet, for
example): a magnetized strip can strongly enhance the critical
current of a superconducting sample and slightly modify the field
structure far from the superconducting strip.

We made all numerical calculations for $d_m=1$. By increasing the
thickness of a magnetic strip it is possible to obtain the same
enhancement of the critical current at a lower magnetization. As
long as $d_m \ll w_m$ we can obtain the same effect if $d_m \cdot
M=const$.

Just as with the magnetic and superconducting screens of different
shapes considered in Refs. \cite{Genenko2,Genenko3}, we believe
that there is an optimal form of a magnetic strip that provides
for a stronger compensation of a current induced magnetic field
than rectangular shape. It is clear that by increasing the
thickness of a magnetic strip at the edges one can enhance the
magnetic field at the edges of a superconducting strip and retain
it in the middle. That allows one to control the current
distribution in the superconducting strip in a more flexible way.

\begin{acknowledgements}

We are grateful to A.A.Fraerman and A.S.Mel'nikov for useful
discussions. The present work was supported by the Russian
Foundation for Basic Research (grant no. 03-02-16774) and by the
Fundamental Program "Quantum Macrophysics" of the Russian Academy
of Sciences. One of us (D.Y.V.) was supported by INTAS Young
Scientist Fellowship 04-83-3139.

\end{acknowledgements}

\end{document}